# Optical emission spectroscopy of non-equilibrium microwave plasma torch sustained by focused radiation of gyrotron at 24 GHz

Sergey Sintsov [1], Kuniyoshi Tabata [2], Dmitry Mansfeld[1], Alexander Vodopyanov[1] and Kimiya Komurasaki[2].

[1] Institute of Applied Physics RAS, 46 Ul'yanov Street, 603950 Nizhny Novgorod, Russia

[2] Department of Aeronautics and Astronautics, The University of Tokyo, 7-3-1 Hongo, Bunkyo-Ku, Tokyo, 113-8656, Japan

E-mail: sins@ipfran.ru

## Abstract

The electron and gas temperature of non-equilibrium argon plasma torch sustained by a focused 24 GHz microwave beam in the external atmosphere of air was measured by optical emission spectroscopy. The excitation temperature of argon atoms was found about 5000 K, while the vibrational and rotational temperatures of ambient nitrogen molecules were respectively estimated at 3000 K and 2000 K. The effective mixing of the external gas atmosphere into the plasma torch is experimentally demonstrated. The vibrational and rotational temperatures of nitrogen molecules slightly increased with the distance from the nozzle exit. Experimental results demonstrate that microwave discharge at atmospheric pressure may be relevant for the decomposition of highly stable molecules, e.g. carbon dioxide, in non-equilibrium conditions.

## 1. Introduction

Over the past few decades, intensive study of plasma jets has led to the development of many successful applications, such as pollutant processing, surface treatment, bacterial inactivation, biomedical treatment, etc. For a number of industrial technologies, such as welding, plasma spraying, melting, waste processing, etc., thermal jets with plasma in local thermal equilibrium (LTE) state, implying approximate equality between heavy particles' and electron's temperature, are widely used. At atmospheric pressure, thermal plasma has electron density up to $10^{15}$ cm$^{-3}$ and high gas temperature (7000 K and above) [1]. However, high gas temperatures, along with high energy consumption and thermal load, are major barriers to a number of applications, such as biomedical treatment, etching surfaces, thin film deposition, etc. That's partly why a lot of attention is paid to cold plasma jets in which plasma is in a non-equilibrium state with gas temperature in the range of 300–1500 K which is at least several times lower than electron temperature [2]. Despite the relatively low plasma density (at atmospheric pressure it usually does not exceed $10^{13}$ cm$^{-3}$ [1]) energetic electrons provide ionization and excitation of heavy particles, generation of radicals and ultraviolet radiation keeping the gas at a relatively low temperature. Studies of such reactive plasma have played an important role in the development of biomedical applications, such as the inactivation of bacteria on heat-sensitive surfaces, wound healing or cancer treatment [3-9]. Cold plasma jets are also successfully implemented in surface etching and thin film deposition [10, 11]. That's partly why a lot of attention is paid to cold plasma jets in which plasma is in non-local-thermodynamic-equilibrium (NLTE), in which the temperature of the heavy particles is in the range of 300–1500 K, while the electron temperature is at least several times lower due to insufficient elastic energy exchange between electrons and heavy particles. There are a number of studies about plasma sources, where the nonequilibrium temperature characteristics are achieved due to the high purge rate of the plasma-forming gas and low heating power [12]. Due to the small energy input, it is impossible to achieve large values of the rates of decomposition of raw gases.

An enormous number of different sources of non-equilibrium plasma jets have been developed at present. Their great variety is explained by the multiple geometries of discharge systems, different operating frequencies as well as the specific use for a particular application. The detailed description and current status of cold atmospheric plasma jets are given in a number of original reviews [2, 13, 14], this article will focus mostly on microwave plasmas. Microwave plasmas are generated and heated by electromagnetic radiation in the range from 300 MHz to 300 GHz). Microwave plasma is versatile since it can be sustained in a wide range of pressures from $10^{-3}$ Pa to $10^5$ Pa. In contrast with other discharges, microwave plasma has higher electron density and higher energy utilization efficiency [14]. Another advantage of microwave plasma is the high life of the discharge system, due to the complete absence of dielectric materials in the discharge area, the surface of which degrades due to dust and bombardment of charged particles.

Atmospheric microwave plasma sources have been studied for decades and many types have been developed, e.g. the microwave continuous flow reactor, surface wave sustained plasma sources (surfatron, surfaguide), the torch with axial gas injection (TIA) and the torch with axial gas injection trough waveguide (TIAGO) design as well as the microwave plasma torch [14, 39-42]. Generally, magnetrons with operating frequencies of 2.4 GHz and 915 MHz and power from tens of watts to several kilowatts are used as microwave sources. The main problems of modern microwave plasma jets are related to the fact that along with an increase of the power absorbed in the discharge, the gas is also heated and the plasma becomes equilibrium. To avoid plasma thermalization, pulsed microwave heating, gases with a high thermal conductivity as well as pressure reduction to subatmospheric values are used, which obviously reduces the attractiveness of the technique for applications. When plasma is heated in a microwave field, an increase in the frequency of the heating radiation allows one to localize and increase the energy input. Potentially, an increase in the energy input leads to an increase in the degree of nonequilibrium temperature characteristics of the plasma and an increase in the plasma torch productivity. Another way to increase the specific energy value is heating in the surface wave mode [18]. In this case, it is possible to achieve non-equilibrium plasma parameters at atmospheric pressure. To organize such a special input of microwave energy, the plasma size should be much smaller than the wavelength. This fact imposes significant restrictions on the amount of power invested in the discharge and its performance. That is why, currently, it seems relevant to study the properties and parameters of a non-equilibrium microwave discharge plasma at a pressure close to atmospheric, which becomes possible with increasing frequency of the microwave field and is associated with the development and implementation of new sources of centimeter and millimeter radiation of high power. Significant progress in the production of technological gyrotron complexes has allowed for the last decade to reach record power levels (up to 20 kW of continuous radiation) with a frequency of ~ 30 GHz with high (up to 60%) efficiency [15]. The idea of using technological gyrotrons with a radiation frequency substantially higher than the traditionally used magnetron frequency of 2.45 GHz to maintain argon plasma torch in non-equilibrium conditions in the external atmosphere of air was recently proposed [16]. One of the advantages of using microwave radiation with a higher frequency is that plasma creation is only possible through the quasi-optical focusing of the microwave beam, which allows the plasma to be removed from the chamber walls or electrodes. Also, at moderate gas pressures, the use of shorter wavelength radiation allows producing more dense plasma, which in combination with a smaller focal spot size leads to an increase in the specific energy density in the plasma. This, in turn, determines the high rate of plasma chemical reactions in such plasma, which makes microwave discharge supported by millimeter waves at atmospheric pressure interesting for a number of applications. For example, a stationary discharge in crossed microwave beams created by powerful radiation with frequencies of 28 GHz with a high plasma density at pressures of about 300 Torr and a high specific energy density of 1-1.5 kW/cm$^{-2}$ was implemented and subsequently successfully applied for diamond CVD growth [17]. Other challenges involve the decomposition of inorganic molecules with high binding energy, e.g. volatile fluorine compounds (SiF$_4$, BF$_3$, GeF$_4$, MoF$_6$), by electron impact [19]. Since fluorine is a monoisotopic chemical element, it is often used in isotope enrichment processes to transfer the enriched substance to the gas phase. After isotope enrichment, it is necessary to deposit the enriched compound in a non-equilibrium plasma. There are a number of papers demonstrating the effective decomposition of volatile fluorides in low-pressure discharges [20-22]. Increasing operating gas pressure up to atmospheric enables not only an increase of the deposition process rate but also avoids the use of expensive corrosion-resistant vacuum equipment. The novel application of non-equilibrium plasma is carbon dioxide decomposition into carbon monoxide to minimize $CO_2$ emission in the atmosphere. The carbon monoxide can be used as a valuable compound for chemistry. This is a possible way to minimize $CO_2$ emission by closed cycle of carbon dioxide. For such process a high electron temperature is also required for step-by-step vibrational excitation of $CO_2$ molecules, while moderate gas temperature avoids vibrational-translational relaxation of excited levels [2].

However, these processes could not be realized in the thermal plasma sources due to the low decomposition rate and low energy efficiency in the equilibrium condition. That's why the implementation of such a non-equilibrium discharge at atmospheric pressure is required for high volume production in the industry.

In this article, we proceed with our previous studies of argon plasma torch [16], supported by the focused continuous microwave radiation of the gyrotron at 24 GHz and power up to 5 kW. The excitation temperature of argon atoms in the plasma torch was measured using optical emission spectroscopy of argon lines. Also, the vibrational and rotational temperatures of nitrogen molecules were analyzed. The degree of mixing of the external gas atmosphere into a plasma torch was estimated. NLTE discharge with a high energy input, supported by focused radiation of the gyrotron, can significantly improve the efficiency of plasma-chemical processes.

## 2. Experimental facilities

### 2.1 Experimental setup

Experiments were carried out using continuous microwave radiation of the gyrotron at 24 GHz. The radiation power can be varied from 0.1 to 5 kW in continuous wave (CW) mode. The gyrotron radiation is the $H_{12}$ mode of a circular waveguide with circular polarization. Microwave radiation from the output of the gyrotron enters the transmission line, where it is transformed into $H_{11}$ mode, and afterward converted into a Gaussian beam. Divergent Gaussian beam electromagnetic radiation is introduced into the gas-discharge chamber through a cooled window made of boron. A parabolic mirror is located inside the camera opposite the radiation input window. The parabolic mirror rotates the microwave beam by 90° and focuses at a distance of 28 cm from its center [16]. The beam cross-sectional square in the waist region is about 1 $cm^2$. The total power transmittance is 85% and the power density of microwave radiation is up to 5 $kW/cm^2$. This corresponds to the intensity of the root mean square (RMS) electric field of 1.9 kV/cm.

The experimental setup is shown in Figure 1. Argon was used as the plasma-forming gas. It was fed into the chamber through a metal tube with an outer diameter of 4 mm. The end of this tube was led to the beam waist region. In this region, a gas discharge was initiated in an argon stream.

The discharge is an elongated torch which length is proportional to the absorbed microwave power and reaches 4 cm. The gas discharge chamber was not sealed from the external atmosphere; therefore, an argon plasma torch was maintained in the air at atmospheric pressure. This method of introducing gas allows you to conveniently mix the reacting components inside the torch. The discharge was stably maintained in continuous mode with argon flow rate from 5 to 30 l/min [16]. A larger argon flow is stabilized by a higher microwave output. Accordingly, for small values of argon flow, it is impossible to raise the power above a certain threshold. In this case, the plasma torch is unstable and tries to escape towards the incident microwave beam.

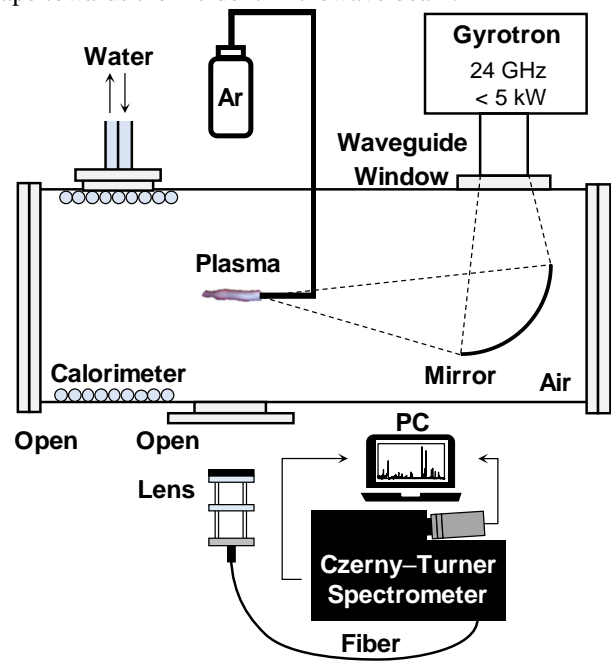

**Figure 1.** Experimental setup and spectroscopic measurement scheme.

Since the transverse size of the plasma torch is smaller than the beam size in the waist region, part of the microwave power was not absorbed by the discharge. The unabsorbed part of microwave power in the plasma torch was measured using a flowing water calorimeter. Its water circuit was located inside the discharge chamber. The part of absorbed microwave power was about 10% of the input power. This roughly corresponds to the ratio between the cross-sectional square of the plasma torch and the cross-sectional square of the microwave beam in the waist region. A number of functional and diagnostic vacuum inputs are located in the gas discharge chamber. It was possible to connect several independent gas inlets and pumps. The gas discharge chamber has quartz glass windows for monitoring the discharge and conducting spectral measurements.

## 2.2 Optical emission spectroscopy

In this work, emission spectra of a plasma torch were obtained. A Czerny–Turner type spectroscope (MS5204i; SOL instruments) with a CCD camera (Digital Camera HS 102H, 2048 pixels resolution; SOL instruments) was employed to record emission spectra in the range of 300–1000 nm with a resolution of 0.028 nm. Figure 1 shows the experimental setup for optical emission spectroscopy. The defocused emission from the plasma torch was collected by a fiber using two lenses. For this experiment, the following spectra in two wavelength regions were obtained because of the different intensities of the spectral lines:

lamp was measured. Then, by comparing the intensity difference between the theoretical intensity of black body radiation of 2999 K and measured intensity, the calibration coefficient was calculated and used to calibrate the intensity. During the procedure, a background line was subtracted from the calibrated spectrum. Furthermore, wavelength calibration was conducted. The calibration was performed by using the peak value of the second positive system (SPS): the vibrational transition (0,2) at 380.63 nm; (0,3) at 406.09 nm.

## 2.3 Measurement of excitation temperature by Ar I lines

The excitation temperature of argon atoms was measured using the Boltzmann plot method by Ar I lines. As the following equation suggests, the excitation temperature can be measured from the inclination in a Boltzmann plot. [16, 38]

Table 1. Spectral data for argon atoms. [37] $\lambda_{i,j}$, $A_{i,j}$, $g_{i,j}$, and $E_i$ are respectively the wavelength, transition probability, statistical weight of upper states, and energy of upper states.

| $\lambda_{i,j}$ /nm | $A_{i,j}$ /s$^{-1}$ | $g_{i,j}$ | $E_i$ /eV | $\lambda_{i,j}$ /nm | $A_{i,j}$ /s$^{-1}$ | $g_{i,j}$ | $E_i$ /eV |
|---|---|---|---|---|---|---|---|
| 415.86 | $1.4 \times 10^6$ | 5 | 14.53 | 696.54 | $6.4 \times 10^6$ | 3 | 13.32 |
| 416.42 | $2.9 \times 10^5$ | 3 | 14.52 | 706.72 | $3.8 \times 10^6$ | 5 | 13.30 |
| 418.19 | $5.6 \times 10^5$ | 3 | 14.68 | 714.70 | $6.3 \times 10^5$ | 3 | 13.28 |
| 419.07 | $2.8 \times 10^5$ | 5 | 14.50 | 727.29 | $1.8 \times 10^6$ | 3 | 13.32 |
| 419.83 | $2.6 \times 10^6$ | 1 | 14.57 | 738.40 | $8.5 \times 10^6$ | 5 | 13.30 |
| 420.07 | $9.7 \times 10^5$ | 7 | 14.50 | 750.38 | $4.5 \times 10^7$ | 1 | 13.48 |
| 425.94 | $4.0 \times 10^6$ | 1 | 14.73 | 751.46 | $4.0 \times 10^7$ | 1 | 13.27 |
| 426.63 | $3.1 \times 10^5$ | 5 | 14.53 | 772.38 | $2.5 \times 10^7$ | 3 | 13.15 |
| 427.22 | $8.0 \times 10^5$ | 3 | 14.52 | 794.82 | $5.2 \times 10^6$ | 3 | 13.28 |
| 430.01 | $3.8 \times 10^5$ | 5 | 14.50 | 800.62 | $4.9 \times 10^7$ | 5 | 13.17 |
| 433.36 | $5.7 \times 10^5$ | 5 | 14.68 | 801.48 | $9.3 \times 10^6$ | 5 | 13.09 |
| 433.53 | $3.9 \times 10^5$ | 3 | 14.68 | 810.37 | $2.5 \times 10^7$ | 3 | 13.15 |
| | | | | 826.47 | $1.5 \times 10^7$ | 3 | 13.32 |
| | | | | 840.82 | $2.2 \times 10^7$ | 5 | 13.30 |
| | | | | 842.46 | $2.2 \times 10^7$ | 5 | 13.09 |
| | | | | 852.14 | $1.4 \times 10^7$ | 3 | 13.28 |
| | | | | 866.79 | $2.4 \times 10^6$ | 3 | 13.15 |

1. 300–900 nm, exposure time 500 ms, which is for the measurement of excitation temperature using 410–440 nm and the measurement of vibrational and rotational temperatures using 370–410 nm; 2. 500–900 nm, exposure time 5 ms or 1 ms, which is for the measurement of excitation temperature using 680–880 nm. The slit width for the experiments 1. and 2. was respectively set as 60 μm and 10 μm. The corresponding full width half maximum (FWHM) of instrumental broadening is 0.12 nm and 0.10 nm, which were measured as the broadening of a neon lamp.

It is necessary to conduct intensity calibration, because the whole optical system including a pair of lenses has a different intensity depending on wavelength and camera's sensitivity also varies depending on its pixels. For this calibration, a calibration lamp was employed, and the intensity from the

$$\ln\left(\frac{I_{i,j} \cdot \lambda_{i,j}}{A_{i,j} \cdot g_i}\right) = -\frac{E_i}{k_B T_{ex}} + C \qquad (1)$$

Therein, $C$ is a constant, and $I_{i,j}$, $\lambda_{i,j}$, $A_{i,j}$, $g_{i,j}$, $E_i$, $k_B$, and $T_{ex}$ are respectively the intensity, the wavelength, transition probability of each Ar I line, statistical weight of upper states, energy of upper states, Boltzmann's constant, and excitation temperature. The constants for the employed Ar I lines are listed at Table 1. In an earlier report [16], the transitions of Ar I excited states that were employed for temperature measurements were limited to the lines from 750 nm to 850 nm. However, Ar I emission lines in a wider range of wavelengths are applicable for more accurate measurement of excitation temperature. Therefore, in this study, emission in 410–440 nm and 680–880 nm was used to measure electron temperature. Most of the lines around 14.5 eV appear in the wavelength region of 410–440 nm, while those around 13.0 eV appear in 680–880 nm. The intensity of the upper states of around 13.0 eV is weaker than that around 14.5 eV, so that the spectra for each wavelength region was measured using a different exposure time as the experiments are respectively denoted as 1. and 2. in the sub-section 2.2. In order to use all the Ar I lines for Boltzmann plot analysis, the measured intensity was calibrated as follows. First, using the intensities at Ar I 714.70 nm measured by the experiments 1. and 2., the intensity ratio of 1. to 2. was calculated. The same procedure was carried out for the line of Ar I 866.79 nm. And then, the average of those values was calculated and employed to calibrate the intensity, so that it became possible to compare the intensities between the Ar I lines that appeared in 410–440 nm and 680–880 nm.

## 2.4 Measurement of vibrational and rotational temperatures by the second positive system

The second positive system (SPS) is often used to measure vibrational and rotational temperatures in air discharge plasma [24, 25]. The spectrum is emitted as transitions from an electronic state C $^3\Pi_u$ to an electronic state B $^3\Pi_g$ and the transitions appear in the ultraviolet region: 300 nm–450 nm. In atmospheric pressure, it takes only 10 ns to sufficiently exchange energy between rotational and translational modes because several collisions are enough for that reaction [26]. Therefore, the rotational temperature is generally regarded as equal to the translational temperature in atmospheric discharge plasma. In the non-equilibrium plasma created by a 24 GHz microwave beam, the emission of nitrogen molecules was observed in 370–382 nm and 392–410 nm. The emission of the first negative system (FNS) by nitrogen molecular ions appeared in 382–392 nm.

The spectrum of SPS can be theoretically calculated by assuming Boltzmann distribution in both vibrational and rotational energy states. In this study, vibrational transitions of the second positive system $\Delta v = -2$ and $-3$ were employed for determining vibrational and rotational temperatures, which respectively appear in the wavelength regions: 370–382 nm and 392–410 nm. According to an earlier report [24], the wavelength that those transitions appear can be determined by the energy difference between electronic states C $^3\Pi_u$ ($v'$, $J'$) and B $^3\Pi_g$ ($v''$, $J''$). Here, $v$ and $J$ respectively denote vibrational and rotational quantum numbers. The energy at each state is the sum of electronic energy $E_e$, vibrational energy $E_{vib}$, and rotational energy $E_{rot}$ as follows.

$$E(v,J) = E_e + E_{vib}(v) + E_{rot}(v,J) \quad (2)$$

Therein, $E_{vib}$ and $E_{rot}$ are calculable by the following equations.

$$E_{vib}(v) = \omega_e\left(v+\frac{1}{2}\right) - \omega_e\chi_e\left(v+\frac{1}{2}\right)^2 \quad (3)$$

$$E_{rot}(v,J) = B_v J(J+1) \quad (4)$$

$$B_v = B_e - \alpha_e\left(v+\frac{1}{2}\right) \quad (5)$$

The coefficients in the equations are listed in Table 2. Consequently, the wavelength $\lambda$ is calculated by the equation (6).

$$\lambda_{Bv'',J''}^{Cv',J'} = \frac{hc}{E_C(v',J') - E_B(v'',J'')} \quad (6)$$

The intensity of the spectrum is expressed by the following equation.

$$I_{Bv'',J''}^{Cv',J'} = \frac{K}{\lambda^4} q_{v'v''} \exp\left(-\frac{E_{vib}}{k_B T_{vib}}\right) S_{J',J''} \exp\left(-\frac{E_{rot}}{k_B T_{rot}}\right) \quad (7)$$

$K$ is a constant. $q_{v'v''}$, $S_{J'J''}$, and $k_B$ respectively denote the Franck–Condon factor, Hönl–London factor and Boltzmann's constant. The Franck–Condon factor corresponding to each vibrational transition was cited from an earlier report [27] as shown in Table 2. In addition, the Hönl–London factors for P, Q, R branches were calculated by the following equations [28].

$$S_P = 6(J+1) - 10/(J+1) \quad (8)$$

$$S_Q = 10/J + 10/(J+1) \quad (9)$$

$$S_R = 6J - 10/J \quad (10)$$

For the experiment to measure vibrational and rotational temperatures, the slit width was set as 60 μm. Therefore, the line shape was given by a gaussian shape with FWHM of 0.12 nm and the spectrum for each vibrational and rotational transition was calculated. Finally, the whole spectrum is calculable by the sum of the broadened spectrum. The least square procedure was employed for deducing vibrational and rotational temperatures. A $\chi^2$ value is defined as a squared difference between measured and calculated values. First, $\chi^2$ value was minimized by changing vibrational and rotational temperatures at an interval of 50 K for each wavelength region 370–382 nm and 392–410 nm. Finally, vibrational and rotational temperatures were calculated as the average values of these temperatures.

Table 2. Coefficients for the second positive system. [24]

|  | N$_2$ (C $^3\Pi_u$) | N$_2$ (B $^3\Pi_g$) |
|---|---|---|
| $E_{ex}$ [$10^4$ cm$^{-1}$] | 8.913 688 | 5.961 935 |
| $\omega_e$ [$10^3$ cm$^{-1}$] | 2.047 17 | 1.733 39 |
| $\omega_e\chi_e$ [$10^1$ cm$^{-1}$] | 2.844 5 | 1.412 2 |
| $B_e$ [cm$^{-1}$] | 1.824 7 | 1.637 4 |
| $\alpha_e$ [$10^{-2}$ cm$^{-1}$] | 1.868 | 1.791 |

Table 3. Franck–Condon factors for vibrational transitions. [27]

| $v'$ | $v''$ | Wavelength /nm | Franck–Condon factor |
|---|---|---|---|
| 0 | 2 | 380.4 | $1.45 \times 10^{-1}$ |
| 1 | 3 | 375.4 | $1.98 \times 10^{-1}$ |
| 2 | 4 | 370.9 | $1.61 \times 10^{-1}$ |
| 0 | 3 | 405.8 | $5.12 \times 10^{-2}$ |
| 1 | 4 | 399.7 | $1.10 \times 10^{-1}$ |
| 2 | 5 | 394.2 | $1.39 \times 10^{-1}$ |
| 3 | 6 | 389.4 | $1.31 \times 10^{-1}$ |

## 3. Results and discussion: dependence on mass flow rate and absorbed microwave power

### 3.1 Excitation temperature

Examples of observed spectra in 410–440 nm and 680–880 nm are shown in figure 2 at the mass flow rate of 15 l/min and absorbed microwave power of 470 W. The spectrum intensity was unchanged for multiple measurements so that the spectrum was recorded once for each condition.

The Boltzmann plot is presented in figure 3. The use of relatively long distance between energy levels of argon atom's upper states enabled to measure the excitation temperature more accurately, though an earlier report used the lines only around 13 eV.

The measured excitation temperature is shown in figure 4. The error bar shown in figure 4 represents the standard error of the approximate line's inclination in figure 3. It should be noted that in the case where an error is calculated in each wavelength region, the standard error will be much larger than the one presented in figure 4. Furthermore, the error that occurs by the intensity calibration using 714.70 nm and 866.79 nm is smaller than the error shown in figure 4.

The results show that there is little change of excitation temperature on the absorbed microwave power and mass flow rate. Measured excitation temperatures were lower than 1–1.5 eV shown in the earlier report [16]. The main reason for the difference is that in the earlier report, the range of measured wavelength region was small, resulting in a large error. The reason why the excitation temperature in the case of 5 l/min and 470 W couldn't be obtained is that a stationary plasma was not maintained. The possibility exists that from the discharge ignition point, an ionization wave propagated toward a microwave source, which phenomenon has been studied for applications such as propulsion [29–33] and ultraviolet light source [34, 35].

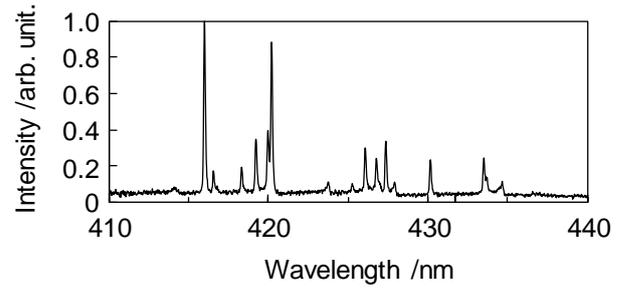
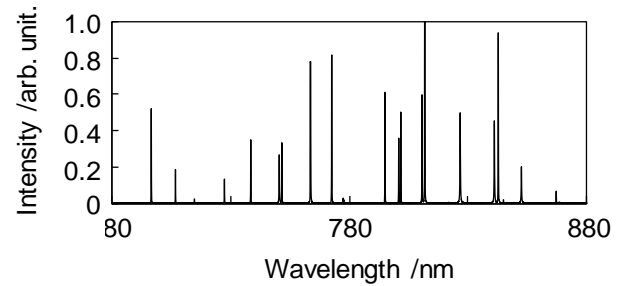

**Figure 2.** Examples of Ar I spectra in 410–440 nm and 680–880 nm at the mass flow rate of 15 L/min and absorbed microwave power of 470 W.

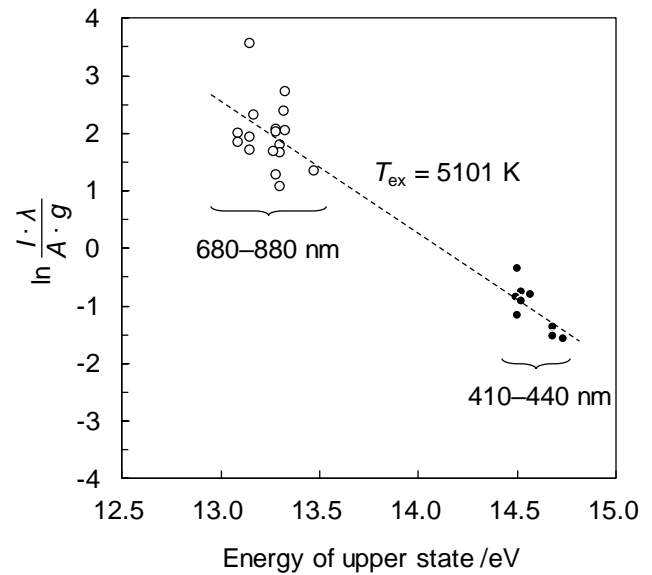

**Figure 3.** An example of a Boltzmann plot at the mass flow rate of 15 L/min and absorbed microwave power of 470 W.

## 3.2 Vibrational and rotational temperatures

An example of observed spectra in 370–410 nm is presented in figure 5. The spectrum intensity was unchanged for multiple measurements so that the spectrum was recorded once for each condition. Measured vibrational and rotational temperatures are shown in figure 4. Each plotted temperature is the average value obtained using the emission in 370–382 nm and 392–410 nm. The spectra's sensitivity to vibrational and rotational temperatures was examined by changing each temperature manually in a theoretical calculation. The results show that the error was at least within ± 300 K. Therefore, the error bar of each plot in the figure 4 respectively stands for the standard error of weighted average, ± 212 K. As an example, the calculated line shapes are presented in figure 6 and 7 for each wavelength region at the mass flow rate of 15 l/min and absorbed microwave power of 470 W.

The results show that the vibrational and rotational temperatures exhibited the same dependence on mass flow rate and absorbed microwave power as the excitation temperature. The reason why the vibrational and rotational temperatures showed the same trend as the excitation temperature is attributed to a similar reason, as mentioned in the section 3.1.

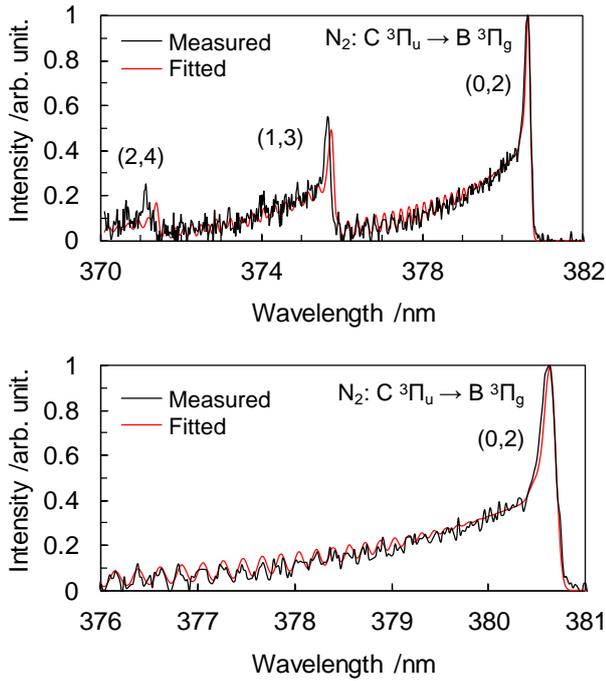

**Figure 6.** Measured and calculated spectra in 370-382 nm at a mass flow rate of 15 l/min and absorbed microwave power of 470 W. $T_{vib} = 2500$ K and $T_{rot} = 1350$ K.

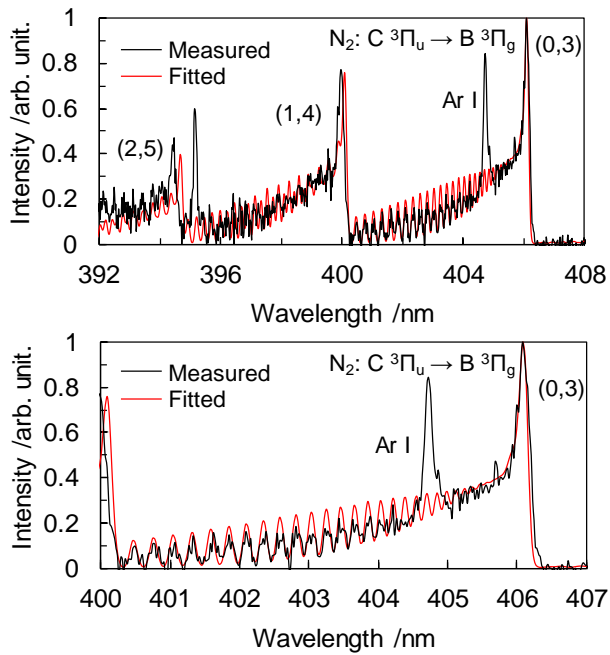

**Figure 7.** Measured and calculated spectra in 392–410 nm at a mass flow rate of 15 l/min and absorbed microwave power of 470 W. $T_{vib} = 2500$ K and $T_{rot} = 1450$ K.

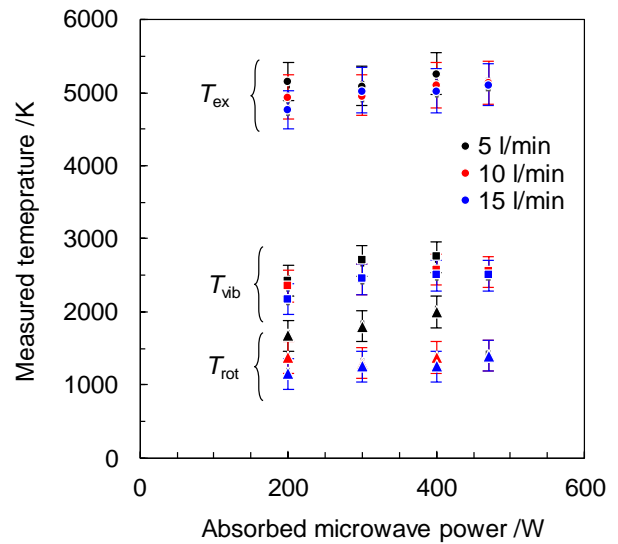

**Figure 4.** Measured excitation, vibrational and rotational temperatures.

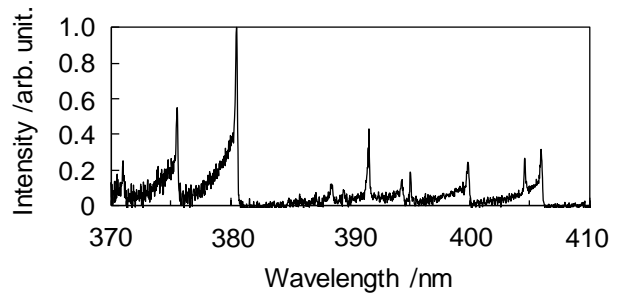

**Figure 5.** A measured spectrum at a mass flow rate of 15 l/min and absorbed microwave power of 470 W.

Translational modes of Ar I and Ar II are considered to be in an equilibrium state with nitrogen molecules' translational mode because of frequent collisions in the boundary region between the argon plasma torch and external gas atmosphere. Therefore, in this study, the gas temperature of argon plasma is regarded as equal to the translational temperature of nitrogen molecules (that is the same as rotational temperature) [26]. Figure 4 shows the measured excitation, vibrational and rotational temperatures for various mass flow rates and absorbed microwave powers. The excitation temperature was 3 times higher than the rotational, and was two times higher, than the vibrational temperature. It was elucidated that the atmospheric discharge plasma supported by focused microwave radiation from a 24 GHz gyrotron is in a non-equilibrium state.

## 4. Results and discussion: one-dimensional structure

### 4.1 Distribution of excitation temperature, and vibrational and rotational temperatures

Figure 8 shows observed plasma structures for absorbed microwave power of 200, 300, and 400 W. As the absorbed power increases, the length of plasma torch increased from 17 mm to 23 mm, which tendency was observed in the earlier report [16]. For the experiments shown in the section 3.1, the temperatures were measured with an enlarged spot diameter to take the whole luminescence of the plasma torch. The figure 8 shows that with an increase of absorbed microwave power, the luminescence of the plasma torch gets stronger and the plasma torch became longer. On the other hand, the measured temperatures didn't show a marked dependence on the absorbed microwave power. Because the emission intensity depends on the number density of corresponding luminescent particles and the excitation temperature, the following things are suggested. The number density of luminescent particles increased with the increase of the microwave energy, and the particles took more time to be quenched, resulting in the growing length of the plasma torch. From the discharge colours in the photographs, the emission of nitrogen molecules was confirmed around the argon plasma torch. The emission of nitrogen molecules was strongest at the absorbed microwave power of 400 W.

Using a system of optical lenses, it was possible to observe emission spectra from the spot diameter of 4 mm from the different location of the torch: (a) 3 mm; (b) 9 mm; (c) 15 mm. For the experiments, integrated luminescence inside the spot region can be obtained as intensity. However, the intensity is written by an exponential function of excitation temperature, so that it is considered that the temperature of the position where the luminescence is the strongest inside the spot region was measured in the experiments. Measured excitation temperature is shown in figure 9. The error bar represents the error of intensity calibration using 714.70 nm and 866.79 nm because it was larger than the standard error of the approximated line's inclination. It cannot be said that the excitation temperature had some dependence on the absorbed microwave power.

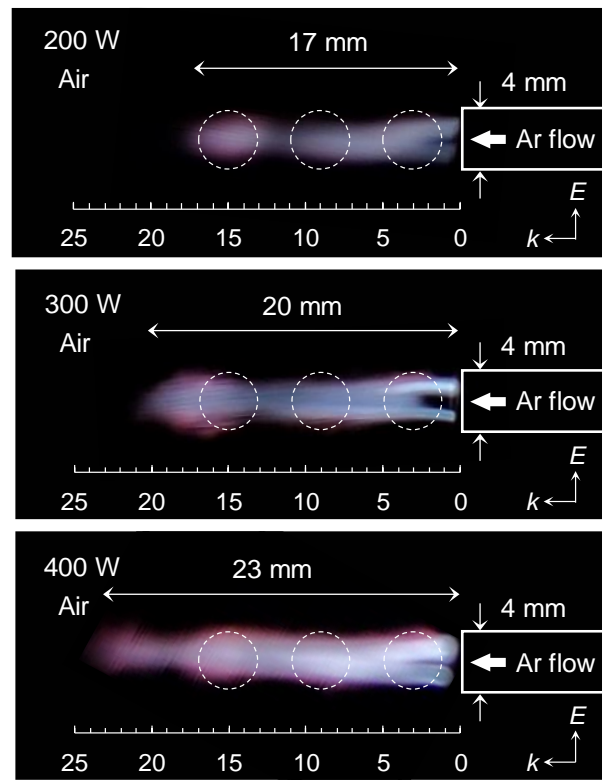

**Figure 8.** Dependence of plasma structure on absorbed microwave power. The exposure time was 1 ms. The location where the emission was analyzed is (a) 3 mm, (b) 9 mm, and (c) 15 mm from the nozzle exit.

Figure 10 shows the one-dimensional distribution of vibrational and rotational temperatures of nitrogen molecules. Both temperatures gradually increased with the distance from the nozzle exit at the absorbed microwave power of 300 W and 400 W. The reason might be attributed to the transportation of electron's kinetic energy to heavy particle's energy. Vibrational and rotational temperatures, as well as excitation temperature, decreased at 9 mm at the absorbed microwave power of 200 W. The possibility exists that the emission was obtained in the region where the luminescence was not so strong as different regions. From the photograph in the figure 8 at 200 W, the emission around 9 mm is weaker than the different regions.

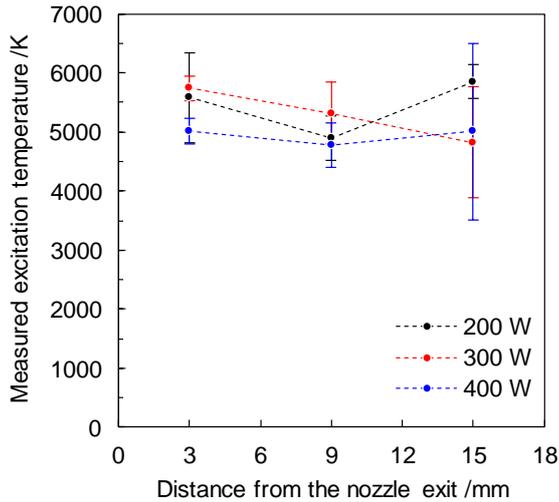

**Figure 9.** One-dimensional distribution of excitation temperature.

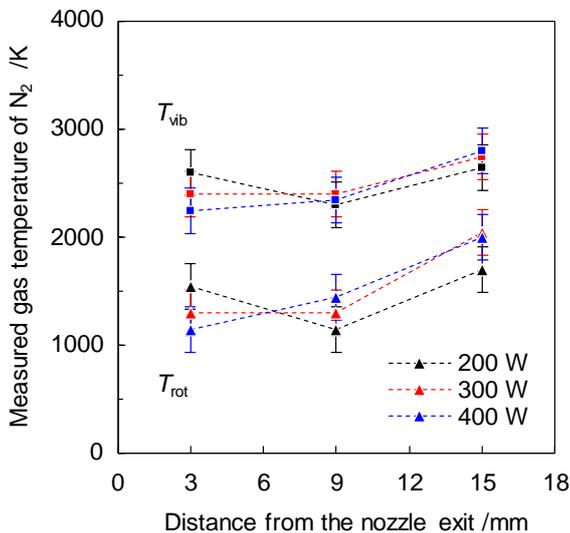

**Figure 10.** One-dimensional distribution of vibrational and rotational temperatures.

*4.2 The number density ratio of nitrogen molecules to Ar I*

To investigate the gas mixing mechanism of the external atmosphere into the argon plasma torch, the number density ratio of nitrogen molecules to Ar I was estimated using the intensity ratio. The spectra at each location (a) 3 mm, (b) 9 mm, and (c) 15 mm from the nozzle exit in 370–440 nm are presented in figure 11.
Firstly, the method to determine the number density ratio is explained. The intensity of the second positive system, $I_{SPS}$ is expressed by equation (11) with the number density of ground state, $[N_2(X)]$. The intensity of Ar I lines: $I_{Ar\,I}$ is written as equation (12) using $i$, which stands for the upper energy state of its emission.

$$\begin{aligned} I_{SPS} &= h\nu \cdot A \cdot [N_2(C)] \\ &= h\nu \cdot A \cdot \frac{[N_2(X)]}{Z_{N_2}} g_{N_2} \exp\left(-\frac{\varepsilon_C}{k_B T_e}\right) \end{aligned} \quad (11)$$

$$I_{Ar\,I} = h\nu \cdot A \cdot \frac{[Ar\,I]}{Z_{Ar}} g_{Ar(i)} \exp\left(-\frac{\varepsilon_{Ar(i)}}{k_B T_e}\right) \quad (12)$$

Therein, $h$, $\nu$, $A$, $g$, $k_B$, and $Z$ respectively denote Planck's constant, frequency of emitted light, transition probability, statistical weight, Boltzmann's constant and partition function. Now, we consider the vibrational transition from the state $v'$ to $v''$ in the second positive system and the electronic transition of Ar I from the state $i$ to $j$. The constants in the equations (11) and (12) other than the number density of Ar I and nitrogen molecules of ground states are respectively written as $C_{N2}(v',v'')$ and $C_{Ar}(i,j)$. Then, the number density ratio of nitrogen molecules to Ar I are expressed by the following equation (13).

$$\frac{[N_2(X)]}{[Ar\,I]} = \frac{C_{Ar}(i,j)}{C_{N_2}(v',v'')} \frac{I_{SPS}}{I_{Ar\,I}} \quad (13)$$

The value $C_{Ar}(i,j) / C_{N2}(v',v'')$ is not credible because the transition probability of the second positive system has some errors. Therefore, the number density ratio for each location was normalized by the value at the location (a) 3 mm, as presented in equations (14) and (15).

$$\left.\frac{[N_2(X)]}{[Ar\,I]}\right|_{(b)} \bigg/ \left.\frac{[N_2(X)]}{[Ar\,I]}\right|_{(a)} = \alpha \quad (14)$$

$$\left.\frac{[N_2(X)]}{[Ar\,I]}\right|_{(c)} \bigg/ \left.\frac{[N_2(X)]}{[Ar\,I]}\right|_{(a)} = \beta \quad (15)$$

Because 11 transitions of electronic states for Ar I and 2 transitions of vibrational states are taken into account, there are totally 22 combinations. Each plot in the figure 12 shows the averaged value. The error bar represents the standard error. The figure 12 suggests that the number density of nitrogen molecules gradually increased with the distance from the nozzle exit. Furthermore, the higher the absorbed microwave power was, more rapidly gas mixing occurred. The luminescence of pink colour corresponds to the second positive system of nitrogen molecules. Therefore, the photographs in the figure 8 also suggests that larger absorbed microwave power enabled to get more rapid gas mixing of the external nitrogen molecules into the argon plasma torch. When the number density ratio was calculated using the equations (11)–(13), excitation temperature was assumed to be constant at any location, so that the number density ratio of nitrogen molecules to Ar I corresponds to the intensity ratio of second positive system to Ar I. It should be noted that the change of excitation temperature significantly affects the intensity ratio since the population in an excited state is expressed by an exponential function with excitation temperature.

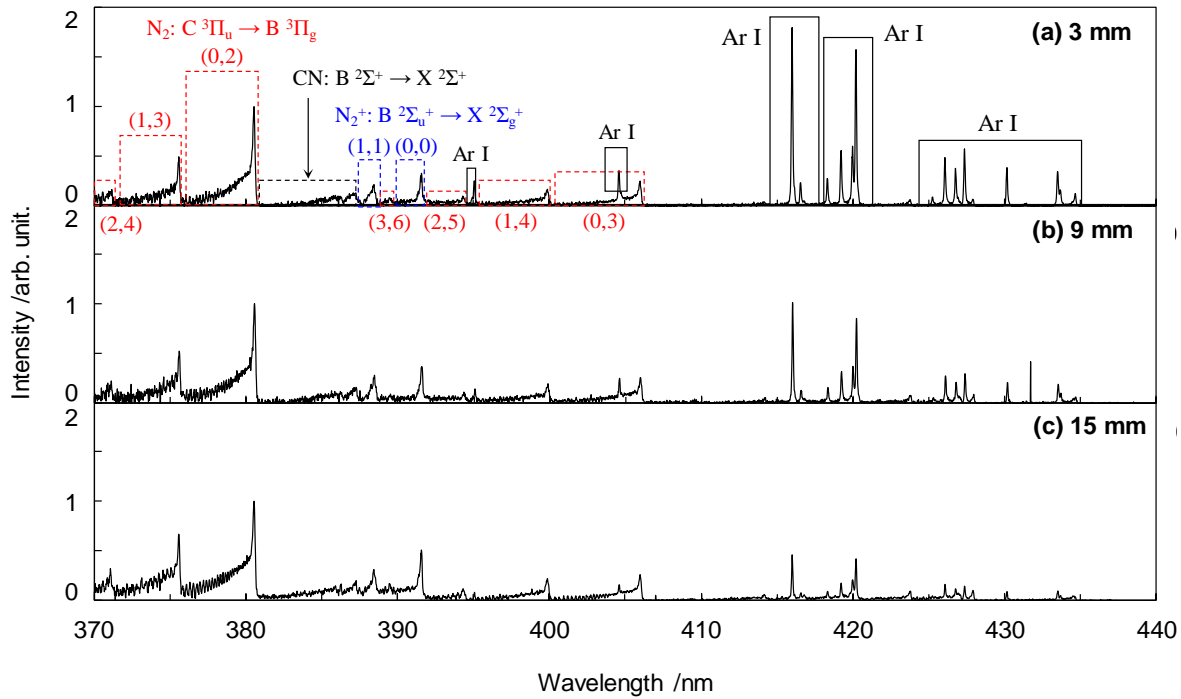

**Figure 11.** Measured spectra at (a) 3 mm, (b) 9 mm, and (c) 15 mm from the nozzle exit.

The result shows that the composition of the plasma-forming mixture along the torch's length varied.

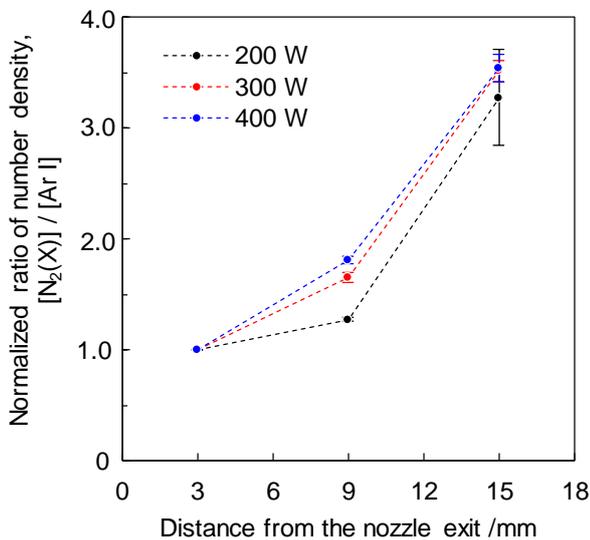

**Figure 12.** One-dimensional distribution of the number density ratio of nitrogen molecules to Ar I, [$N_2$(X)] / [Ar I]. The values at 9 mm and 15 mm respectively show $\alpha$ and $\beta$.

In particular, the concentration of nitrogen molecules from the external atmosphere increases by about 3 times towards the end of the torch. The use of argon as a plasma-forming gas allows one to obtain a non-equilibrium gas discharge at atmospheric pressure. This result is a characteristic of the effectiveness of this type of organization of plasma-chemical processes, where a non-equilibrium argon torch reacts with the surrounding gas atmosphere.

## 5. Conclusion

The electron and gas temperature of non-equilibrium discharge plasma sustained by a continuous 24 GHz microwave beam, was measured by optical emission spectroscopy. The excitation temperature of argon atoms was found about 5000 K, while the vibrational and rotational temperatures of ambient nitrogen molecules were respectively estimated at 3000 K and 2000 K, which slightly depends on the mass flow rate of argon and absorbed microwave power. Furthermore, the one-dimensional structure of the plasma torch was also investigated. It turned out that the external atmosphere was gradually mixed into the argon plasma torch and the vibrational and rotational temperatures of nitrogen molecules slightly increased with the distance from the nozzle exit. The effective mixing of the external gas atmosphere into the plasma torch was experimentally demonstrated.

One of the possible applications of non-equilibrium plasma in proposed microwave discharge is carbon dioxide decomposition into carbon monoxide, in order to minimize $CO_2$ emission in the atmosphere. The excitation temperature of 0.5–1 eV is enough for step-by-step vibrational excitation of $CO_2$ molecules, while moderate gas temperature avoids vibrational-translational relaxation of excited levels [36]. Using carbon dioxide instead of air as an external atmosphere will allow to make its effective conversion into carbon monoxide in the discharge supported in non-equilibrium conditions by powerful microwave radiation of gyrotron at atmospheric pressure.

## Acknowledgements

The work was supported by RFBR grant #18-48-520027 and JSPS KAKENHI Grant Number JP15H05770.